\documentstyle[12pt]{article}
\catcode`@=11   
\def\NPB{{\it Nucl.~Phys.~}{\bf B}}
\def\PLB{{\it Phys.~Lett.~}{\bf B}}
\def\PRL{{\it Phys.~Rev.~Lett.~}}
\def\PRD{{\it Phys.~Rev.~}{\bf D}}

\def\IJMPA{{\it Int.~J.~Mod.~Phys.~}{\bf A}}

\let\a=\alpha
\def\tA{\tilde{A}}

\def\tB{\tilde{B}}

\def\rd{{\rm d}}
\def\define{\buildrel{\rm def}\over=}
\let\f=\phi

\def\inv#1{{\textstyle{1\over#1}}}

\let\L=\Lambda
\def\Lb{\Lambda_b}
\let\q=\theta

\let\p=\pi
\def\Seff{S_{\rm eff}}
\let\t=\tau
\let\vd=\partial
\def\vev#1{\left\langle#1\right\rangle}

\def\tZ{\tilde{Z}}
\let\w=\omega

\def\IR{\relax\leavevmode{\rm I\kern-.18em R}}
\def\ZZ{\relax\leavevmode
       \ifmmode\mathchoice
       {\hbox{\sf Z\kern-.4em Z}}
       {\hbox{\sf Z\kern-.4em Z}}
       {\lower.9pt\hbox{\scriptsize\sf Z\kern-.36em Z}}
       {\lower1.2pt\hbox{\tiny\sf Z\kern-.36em Z}}
       \else{\sf Z\kern-.4em Z}\fi}
\def\RR{\relax\leavevmode
       \ifmmode\mathchoice
       {\hbox{\sf R\kern-.4em R}}
       {\hbox{\sf R\kern-.4em R}}
       {\lower.9pt\hbox{\scriptsize\sf R\kern-.36em R}}
       {\lower1.2pt\hbox{\tiny\sf R\kern-.36em R}}
       \else{\sf R\kern-.4em R}\fi}

\def\resetby#1#2{\@addtoreset{#2}{#1}}
\def\seceq{\@addtoreset{equation}{section}
              \def\theequation{\thesection.\arabic{equation}}}

\def\Label#1{\label{#1}%
                \smash{\hbox to0pt{\raise1ex\hbox{\tiny[#1]}\hss}}}
\def\noLabels{\let\Label=\label}
\def\Eq#1{Eq.~(\ref{#1})}

\thicklines     

\reversemarginpar
\catcode`@=12
\begin{document}

\begin{titlepage}
\begin{flushright}
CITUSC/02-03\\
VPI-IPPAP-02-01\\
hep-th/0201187\\
\end{flushright}

\begin{center}

{\large\bf
           {Relating the Cosmological Constant and Supersymmetry Breaking
            in Warped Compactifications of IIB String Theory}}\\[5mm]
{\bf P.~Berglund\footnote{e-mail: berglund@citusc.usc.edu} } \\[1mm]
            CIT-USC Center for Theoretical Physics\\
            Department of Physics and Astronomy\\
            University of Southern California\\
            Los Angeles, CA 90089-0484\\[2mm]
{\bf T.~H\"{u}bsch\footnote{e-mail: thubsch@howard.edu}%
            $^,$\footnote{On leave from the ``Rudjer Bo\v skovi\'c''
            Institute, Zagreb, Croatia.} } \\[1mm]
            Department of Physics and Astronomy\\
            Howard University\\
            Washington, DC 20059\\[2mm]
{\bf D.~Minic\footnote{e-mail: dminic@vt.edu} } \\[1mm]
            Institute for Particle Physics and Astrophysics\\
            Department of Physics\\
            Virginia Tech\\
            Blacksburg, VA 24061\\[5mm]

{\bf ABSTRACT}\\[3mm]
\parbox{4.9in}{It has been suggested that the
observed value of the cosmological constant is
related to the supersymmetry breaking scale $M_{susy}$
through the formula
$\Lambda \sim M_p^4 (M_{susy}/M_p)^8$.
We point out that a similar relation naturally arises in the
codimension two solutions of warped
space-time varying compactifications of string theory in
which
non-isotropic stringy moduli induce
a small but positive cosmological constant.}
\end{center}
\end{titlepage}

Recently, we addressed the question of whether de Sitter
space~\cite{dsrev} can be obtained from string theory~\cite{bhmds}.
Such non-singular, non-static spacetimes
fall into the class of codimension two non-supersymmetric
string vacua studied in Refs.~\cite{bhm1, bhm3, bhm4}\footnote{By
string vacua we denote solutions that satisfy the corresponding type
IIB supergravity equations of motion and contain moduli with
proper $SL(2,\ZZ)$ properties.}. In these models, supersymmetry
is explicitly broken by a global cosmic brane~\cite{cohen} with a
core of size $\ell$, extended along the $D-2$  ``longitudinal''
directions.
While Refs.~\cite{bhm1, bhm3, bhm4} considered a flat Minkowski
brane, the main point of Ref.~\cite{bhmds} is the
existence of non-supersymmetric codimension two solutions
with a positive cosmological constant, $\Lb$, in the $D-2$
dimensional longitudinal space\footnote{
Note that $\Lb>0$, {\it removes} the naked singularity present in the
model considered in~\cite{bhm1, bhm3, bhm4}, in comparison with earlier
discussions of a positive cosmological constant along the
brane-world~\cite{kachru}.}.

Since the cosmological constant in our model~\cite{bhmds} is
directly related to the non-isotropy of matter, we may compare with
various attempts to
incorporate Mach's principle in string theory~\cite{mach,HM}
as well with the idea that supersymmetry breaking might have
a cosmological origin~\cite{banks,DeoGates}.
In particular, it has been suggested that
the observed value of the cosmological constant,
$\L_4\sim10^{-44}$\,GeV$^4$,~\cite{cosmoc,banks1,kachru} may be
related to the supersymmetry breaking scale $M_{susy}$
through the formula~\cite{kiritsis,kim,banks}:
\begin{equation}
    \L_4 \sim M_p^4\> \Big({M_{susy}\over M_p}\Big)^8~,
    \label{e:BanksCC}
\end{equation}
with $M_{susy}\sim10$\,TeV and $M_p\sim10^{19}$\,GeV as
appropriate in 4-dimensions\footnote{The essential ingredient
of this proposal is a conjectured relevance of a
non-decoupling between the microscopic and macroscopic
degrees of freedom \cite{HM, banks, banks1,kachru} for the cosmological
constant problem.
This conjecture is natural from the following intuitive
perspective on the cosmological constant problem. On one hand,
the cosmological constant is tied to the fundamental physics of the
vacuum, because $\L_4$
is essentially given by the vacuum energy density. On the other hand,
the cosmological constant is related to the large scale behavior of
the universe, since a small
cosmological constant implies that the observable universe is big and
almost
flat.}.
It is the aim of this note to point out that the set-up of
Ref.~\cite{bhmds} for $D=6$ leads
naturally to a relation analogous to Eq.~(\ref{e:BanksCC}). In particular, the stringy
moduli induce a non-trivial relation between the scale of the global
cosmic brane,
$\ell$,
the non-isotropy of matter, $\omega$, induced by the brane,
and $\Lb$~\cite{bhmds}.
We take this observation one step further and find an explicit
relation between $\Lb$, $\ell$ and the natural mass scales in this
theory, $M_{6}$ and $M_{4}$,
the Planck scales in the bulk and along the brane, respectively,
thus deriving an equation analogous to Eq.~(\ref{e:BanksCC}).

Although the detailed physics leading to this relation is unclear,
we find it very intriguing that an equation similar to Eq.~(\ref{e:BanksCC})
emerges naturally in our framework.
Still, one of the
most important unresolved questions in
the scenario presented in Ref.~\cite{bhmds} is the issue of the stability
of this non-supersymmetric background in full string theory.
Since supersymmetry is broken, one also has to
address the effects of stringy corrections. We will argue that those
corrections
are negligible.

The general framework of our analysis is as
in Refs.~\cite{bhm1,bhm3,bhm4,bhmds}, to which we refer the reader for
a more
detailed analysis.
Although we will be mostly interested in the phenomenologically relevant case in which $D=6$ and hence the uncompactified spacetime is
$D{-}2=4$-dimensional we find it useful to work in a general
$D$-dimensional background.
We consider Type~IIB
string theory (compactified on a {\it fixed\/} supersymmetry preserving space) in which
the axion-dilaton system, $(\a,\f)$, described by the complex
modulus field $\t=\a + i\exp(-\phi)$,
varies over the $x_{D-2,D-1}$ directions of the uncompactified spacetime.
Thus, the relevant part of the low-energy effective
$D$-dimensional action of the modulus, $\t$,
coupled to gravity reads
\begin{equation}
                \Seff
                = {1\over2\kappa^2}\int\rd^D x \sqrt{-g} ( R
                   - {\cal G}_{\t \bar{\t}}g^{\mu \nu}
                     \vd_{\mu} \t \vd_{\nu} \bar{\t}+\ldots)~.
\label{e:effaction}
\end{equation}
Here $\mu,\nu=0,{\cdots},D-1$, $2\kappa^2=16\p G^{(D)}_N$, where
$G_N^{(D)}$ is the $D$-dimensional Newton constant, and ${\cal
G}_{\t\bar\t}=-(\t-\bar \t)^{-2}$ is the metric on 
the complex structure moduli space of a torus\footnote{Recall that because of its $SL(2,\ZZ)$ properties, the axion-dilaton, $\tau$, can be thought of as the complex structure of a $T^2$, in analogy with F-theory~\cite{Vafa}.}.

Let us now briefly review the codimension two solution with positive
cosmological constant, $\Lb$, along the longitudinal direction of the
cosmic brane~\cite{bhmds}.
The metric Ansatz is:
\begin{eqnarray}
           \rd s^2 &=& A^2(z)\, \bar g_{ab}\rd x^a \rd x^b
                     + \ell^2 B^2(z)\,(\rd z^2 + \rd\q^2)~,
\label{e:Metric} \\[1mm]
           \bar g_{ab}\rd x^a \rd x^b &=& - \rd x_0^2 +
                e^{2\sqrt{\Lb} x_0}\,(\rd x_1^2 + \ldots + \rd
x_{D-3}^2)~,
                     \label{e:gbar}
\end{eqnarray}
where $z=\log(r/\ell)$. As in the case when $\Lb=0$, we find that
the explicit solutions for $\t$ are aperiodic, such as $\t=\alpha_0+i
g_s^{-1}\exp(\omega \q)$, but do exhibit a non-trivial $SL(2,\ZZ)$
monodromy~\cite{bhm1}\footnote{Although $\vd\tau$ does not transform
correctly under $SL(2,\ZZ)$ transformations, it is straightforward to
show that ${\cal G}_{\t \bar{\t}}^{-1}|\vd\tau|^2$, which appears in the
action~(\ref{e:effaction}), is invariant.}.
This ensures our solution to be stringy (although classical and
non-supersymmetric) rather than merely a supergravity vacuum. Note in
particular that the dilaton of the Type~IIB superstring theory varies
with the polar angle, not the radial distance.
 Recall that with $\t=\alpha_0 + i
g_s^{-1}\exp(\omega \q)$, the $SL(2,\ZZ)$ symmetry requires $g^D_s\sim
O(1)$ in $D$ dimensions. However, in the $D{-}2$-dimensional
brane-world, $g_s^{D-2}=g_s^D\sqrt{\a'/V_\perp}$, and since $V_\perp$,
the volume of the transversal space, is  large~\cite{bhm1}, $g_s^{D-2}\ll1$.
Below, we will return to discussing the
corrections to our classical solution.

Following~\cite{bhmds}, the Einstein equation can be simplified
\begin{equation}
         R_{\mu\nu} ~=~ {\cal G}_{\t\bar{\t}}\,\vd_{\mu}
\t\vd_{\nu}\bar\t
         \define\widetilde{T}_{\mu\nu}~.
\label{e:EinStein}
\end{equation}
Since the metric~(\ref{e:Metric}) is
axially symmetric, while
$\t$ is independent of the radial distance from the cosmic brane,
$\widetilde{T}_{\mu\nu}=\hbox{diag}[0,{\cdots},0,\inv4\w^2\ell^{-2}]$.
\Eq{e:EinStein} then defines the general class of our spacetimes as
{\it almost\/} Ricci-flat:
$R_{\mu\nu}=\hbox{diag}[0,{\cdots},0,\inv4\w^2\ell^{-2}]$,
where $\w^2{>}0$ is indeed related to supersymmetry
breaking~\cite{bhm1} and $\ell$ is the (transversal) length scale
of the cosmic brane.

The $R_{ab}=0$ part of \Eq{e:EinStein} reduces
to a single equation, giving:
\begin{equation}
         B^2 =  \ell^{-2}\Lb^{-1}\Big(A'{}^2+{1\over(D{-}3)} A A''\Big)
   =\ell^{-2}\Lb^{-1} \frac{h''  h^{-\frac{D-4}{D-2}}}{(D-2)(D-3)} ~,
\label{e:BfromA}
\end{equation}
which determines $B(z)$ in terms of $A(z)$ or $h(z)\define A(z)^{D-2}$.
With this substitution, the remaining components of
\Eq{e:EinStein} reduce to the following equation:
\begin{equation}
     \frac{1}{2(D-2)}\frac{h'{}^2}{h^2} - \frac{h''}{2h}
+\frac{h'h'''}{2hh''}= -\frac{1}{8}\omega^2\,.~~~~ \label{e:h}
\end{equation}
For $\w\neq0$ ($\t\neq\hbox{\it const.}$), \Eq{e:h} has a
perturbative, analytic solution\footnote{This solution is of the same
form
as that discussed by Gregory~\cite{RG} for the $U(1)$ vortex solution.}:
\begin{eqnarray}
   A(z) &=& Z(z) \Big(1- {\w^2 \rho_0^2(D-3)\over 24
            (D-1)(D-2)} Z(z)^2 + O(\w^4)\Big)~,
\label{e:A}\cr
   B(z) &=& {1\over\ell\rho_0\sqrt{\Lb}}\Big(1 -  {\w^2\rho_0^2\over
             8(D-1)} Z(z)^2 + O(\w^4)\Big)~, \label{e:B}
\end{eqnarray}
where $Z(z)=1-z/\rho_0$ and $\rho_0>0$.
As was shown in Refs.~\cite{bhmds},
close to the horizon spacetime is asymptotically flat in agreement with
the behavior of Rindler space~\cite{kaloper}.

In contrast, when $\Lb=0$ the solution is very different~\cite{cohen,bhm1},
\begin{equation}
           \tA(z) = \tZ(z)^{1\over(D-2)} ~,\qquad
           \tB(z) = \tZ(z)^{-(D-3)\over 2(D-2)}
                   e^{{\xi\over 2a_0}(1-\tZ(z)^2)}~,
\label{e:oldsolution}
\end{equation}
where now $\tZ=(1-a_0 z)$, and we restrict to $a_0>0$. This
solution exhibits a naked singularity, at $z=a_0^{-1}$ ($\tZ=0$),
for the global cosmic brane.

While the naked singularity has been removed by $\Lb>0$,
it was first shown by Gregory~\cite{RG} and by~\cite{bhmds} that the global cosmic brane solution~(\ref{e:oldsolution}) is still a good approximation to \Eq{e:B} away from the horizon.
In particular, by comparing \Eq{e:B} with \Eq{e:oldsolution} close
to the core one can show that
\begin{eqnarray}
a_0 &=&-{h'\over h}|_{z=0}~,\qquad\qquad 
\xi = \Big({h''\over 2 h'} - {\omega^2 h \over 8 h'}\Big)|_{z=0}\label{e:a0-xi}\\
\ell &= & \Lambda_b^{-1/2}\sqrt{ {h'' h^{-{(D-4)\over (D-2)}}\over (D-2)(D-3)}|_{z=0}}~.
\label{e:ell-lambda}
\end{eqnarray}
That is, given a smooth solution defined by~(\ref{e:A}) and parameterized in terms of $(\rho_0,\omega,\Lambda_b)$,  this solution close to $z=0$ can be interpreted as a global cosmic brane solution with parameters $(a_0,\xi,\ell)$ determined by the $(\rho_0,\omega,\Lambda_b)$ through Eqs.~(\ref{e:a0-xi}) and~(\ref{e:ell-lambda}). 
Alternatively, we can solve for $\Lambda_b$,
\begin{equation}
\Lambda_b={\Big(\omega^2 - \omega^2_{GCB} A^2|_{z=0}\Big)\over 4 \ell^{2} (D-2)(D-3)}\define{\Delta \omega^2\over 4 \ell^{2} (D-2)(D-3)}~,
\label{e:dS-omega}
\end{equation}
where $\omega^2_{GCB}\define8 a_0\xi$~\cite{bhm1}. Note that $\omega_{GCB}^2$ is the stress tensor associated to the global cosmic brane to which the solution asymptotes when $z\to 0$, while $\omega^2$ is the stress tensor for the $\Lambda_b>0$ solution. 
Thus, the cosmological constant is directly related to the non-trivial
variation of the matter as a function of $\q$!
This gives a very non-trivial relation between the stringy moduli,
and hence string theory itself, and a positive $\Lb$.
Furthermore, $\Lambda_b>0$ implies that $\omega^2 > \omega_{GCB}^2$.\footnote{That $\Lambda_b$ is indeed positive can be seen from Eq.~(\ref{e:BfromA}). At the horizon, $A(z{=}\rho_0)=0$, which implies that the right hand side of Eq.~(\ref{e:BfromA}) is positive if $\Lambda_b>0$.} When $\omega^2 =0$ it then follows that $\omega^2_{GCB}=0$. The latter is a necessary condition for obtaining a supersymmetric configuration. Thus, we see the important relation between supersymmetry breaking and a positive cosmological constant.

Finally, the Newton constant, $G^{(D-2)}_N=M_{D-2}^{-(D-4)}$, in
$D{-}2$ dimensions and the zero-mode wave function normalization,
$\vev{\psi_0|\psi_0}$, are~\cite{bhmds}:
\begin{equation}
      G^{(D-2)}_N=
     M_{D}^{-(D-2)} \vev{\psi_0|\psi_0}^{-1}~,\qquad\mbox{and}\qquad
\vev{\psi_0|\psi_0}\sim{\pi\over D{-}3}{\ell\over\sqrt{\Lb}}~.
\label{GN}
\end{equation}
The volume of the transversal space, $V_\perp=\vev{\psi_0|\psi_0}$,
is large~\cite{bhmds} and drives the large $M_{D-2}/M_D$ hierarchy.
This then implies the following relation,
\begin{equation}
\L_{D-2}\sim \Big({\pi\over D{-}3}\Big)^2
M_{D-2}^{~D-2}~(\ell\,M_{D-2})^2~\Big({M_D\over M_{D-2}}\Big)^{2D-4}~,
   \label{e:energydensity}
\end{equation}
where $\L_{D-2}=\Lb/G^{(D-2)}_N$ is the energy density in $D{-}2$ dimensions.

{}From now on we will focus on the phenomenologically relevant case of $D=6$.
Recall that $\ell$ is the characteristic (transverse) size of the cosmic
brane, for the formation of which no concrete physical mechanism is
known. However, should $\ell$ be stabilized by a longitudinal
$4$-dimensional physics mechanism~\footnote{There exist both field and string theory 
arguments of this type~\cite{ellM4}.},
then $\ell\sim M_{4}^{-1}$ and (up to factors of ${\cal O}(1)$)
\begin{equation}
   \L_{4}\sim 
   M_{4}^{~4}~\Big({M_6\over M_{4}}\Big)^{8}~.
   \label{e:LongCC}
\end{equation}
The original scenario of Ref.~\cite{bhmds} then applies, where the
10-dimensional spacetime of the Type~IIB string theory is
compactified on a 4-dimensional supersymmetry preserving
space\footnote{All remaining
supersymmetry will be broken by the cosmic brane
solution~\cite{bhm1}.} of characteristic size
$M_{10}^{-1}=M_6^{-1}\sim(10\,\mbox{TeV})^{-1}\sim10^{-19}$\,m. The
cosmic brane of Ref.~\cite{bhmds} then describes a 3+1-dimensional
de~Sitter
world-brane, with the characteristic scale
$M_4\sim10^{19}\,$\,GeV. 
Furthermore, 
$L\define\Lambda_b^{-1/2}\sim 10^{41}\,\mbox{GeV}^{-1}\sim 10^{25}$\,m, provides a natural scale which coincides with the Hubble radius.


Note that Eq.~(\ref{e:LongCC}) is an equation of the same form as
the desired relationship~(\ref{e:BanksCC})
upon identifying $M_{10}=M_6$ with the scale of supersymmetry
breaking, and $M_4$ with the four dimensional Planck scale,
$M_p$.
More precisely Eq.~(\ref{e:LongCC}) provides an explicit relation between
the value of the cosmological constant and the hierarchy involving the two fundamental
scales. Without a detailed dynamical mechanism it is of course very difficult to argue
that $M_6$ should be precisely identified with the scale of supersymmetry breaking.
Nevertheless, as we will indicate in the concluding paragraph, the idea that the cosmological constant
and supersymmetry breaking are related is natural in our model.
As far as we know, this is the first time
such a relation between the observed value of the cosmological
constant and the scale of supersymmetry
breaking has been obtained in a specific dynamical
situation.
Note that this relation crucially depends on the
fact that the zero mode normalization scales as
$\vev{\psi_0|\psi_0}\sim
{\ell\over\sqrt{\Lb}}$, which is a specific feature of the scenario
presented in \cite{bhmds}.

In fact, there exists a whole spectrum of scenarios, albeit with powers
of the mass scale ratio in Eq.(\ref{e:energydensity}) 
which may
not be 8 as in \Eq{e:LongCC}. These scenarios differ in the
compactification/cosmic brane Ansatz sequencing.
For example, let the 10-dimensional spacetime of the Type~IIB string theory
first be compactified on a 3-dimensional supersymmetry preserving space
of characteristic size
$M_{10}^{-1}=M_7^{-1}\sim(10\,\mbox{TeV})^{-1}\sim10^{-19}$\,m.
Assuming that $\ell$ is stabilized by the ``bulk'' $7$-dimensional
physics, then $\ell\sim (M_7)^{-1}$ and
$\L_{5}\sim M_{5}^{~5}~\Big({M_7\over M_{5}}\Big)^{8}$.
Upon a Kaluza-Klein compactification on a circle of radius
$M_5^{-1}\sim(10^{19}\,\mbox{GeV})^{-1}\sim10^{-45}$\,m, this yields
a 3+1-dimensional de~Sitter
world-brane with
$M_4=M_5\sim10^{19}\,$GeV.
On the other hand,
for a codimension two cosmic brane in 10 dimensions
with $\ell$ stabilized by the longitudinal 8-dimensional physics,
$\L_8\sim M_8^8~({M_{10}\over M_8})^{16}$.
After wrapping on a suitable 4-dimensional space (of size $M_8^{~-1}$),
for the desirable values of $\L_4\sim10^{-44}$\,GeV$^4$ and
$M_4=M_8\sim10^{19}$\,GeV, we find that
$M_{10}\sim5.6{\times}10^6$\,GeV is the fundamental scale.
  At the opposite end, by compactifying the 10-dimensional spacetime on a
suitable 4-dimensional space and then constructing a codimension two
cosmic brane in 6 dimensions with $\ell$ stabilized
by the ``bulk'' 6-dimensional physics,
$\L_4\sim M_4^4~({M_6\over M_4})^6$. For the desirable values of
$\L_4\sim10^{-44}$\,GeV$^4$ and $M_8=M_4\sim10^{19}$\,GeV, the fundamental
scale becomes $M_6=M_{10}\sim100$\,MeV.

Finally, let us conclude by discussing the stringy and quantum corrections
to our solution. We will assume that the 6-dimensional theory has
the equivalent of $N=4$ supersymmetry in  4 dimensions, or
equivalently 16 supercharges. This will always be the case as long as
we are considering type II theories with at most a
K3-compactification from ten to six dimensions. First, note that
$\Lb\sim h''|_{z=0}\ell^{-2}$ (which follows from Eq.~(\ref{e:ell-lambda}))
is consistent with the notion that supersymmetry
breaking and a non-zero cosmological constant are related.
To see this, first recall that, from Ref.~\cite{bhm3}, the supersymmetry breaking is
indicated by the non-vanishing of ${\rd A\over\rd r}$. But $A'\sim A h'/h$ so supersymmetry is broken when $h'\neq 0$, which in turn implies that $h''\neq 0$ and hence $\Lambda_b>0$~\footnote{If $h'\neq 0$ and $h''=0$ then this describes the singular global cosmic brane solution in which $\Lambda_b=0$.}.
Furthermore, from $h=A^{D-2}$ it follows that (at least close to the horizon) $A' \sim (h'' h^{-{D-4\over D-2}})^{1/2} \sim \Lambda_b^{1/2} \ell^{-1}$. With $\ell \sim M_{4}^{-1} = 10^{-19}$\,GeV$^{-1}$ and
$\Lambda_b \sim  10^{-82}$\,GeV$^2$ we find $A'\sim 10^{-60}$ which is a very small number. 
This, we argue, justifies neglecting the corrections due to
supersymmetry breaking\footnote{In general there will be $\alpha'$
and string coupling corrections without breaking supersymmetry; these
will not be considered here.}.
The $\alpha'$ corrections due to the global cosmic string would have
to take the form
$\alpha'/V_\perp$ where
$\alpha'\sim M_6^{-2}$ is the string scale.
{}From \Eq{GN} it then follows that the string corrections are very
small.
Now, although our solution is not BPS, supersymmetry is broken very
weakly.
Therefore, the corrections should be proportional both to the coupling and
the
supersymmetry breaking parameter. Since the six-dimensional string coupling $g^6_s\sim O(1)$ because of
modular invariance, the four-dimensional string coupling
$g^{4}_s\ll1$ as discussed above.
Then, the smallness of the
supersymmetry breaking parameter justifies neglecting strong
coupling corrections.

{\bf Acknowledgments:}
We are indebted to V.~Balasubramanian, J.~de~Boer, S.~Kachru, N.~Kaloper
and F.~Larsen
for very useful discussions.
P.~B. would like to thank the organizers of the Lake Arrowhead
workshop on ``New Directions in Conformal Field Theory'' for a
stimulating environment, as well as CIG, Berkeley.
    T.~H.\ wishes to thank the Caltech-USC Center for
Theoretical Physics for its hospitality.
     The work of P.~B.\ and T.H.\ was supported by the US
Department of Energy under grant numbers DE-FG03-84ER40168 and
DE-FG02-94ER-40854, respectively.
    D.~M. would like to thank the organizers of the KIAS
Winter School in String Theory for hospitality and for providing a
stimulating working environment.

\end{document}